\documentclass[12pt]{iopart}
\usepackage{iopams}  
\usepackage{graphicx}
\begin{document}


\title[comparison of searches for bursts using LIGO and Virgo simulated data  ]
      {A first comparison of search methods for gravitational wave bursts using LIGO and Virgo simulated data }

\author{ L.~Blackburn$^3$,
F.~Beauville$^5$,
M.-A.~Bizouard$^7$,
L.~Bosi$^8$,
P.~Brady$^4$,
L.~Brocco$^9$,
D.~Brown$^{2,4}$,
D.~Buskulic$^5$,
S.~Chatterji$^2$,
N.~Christensen$^1$,
A.-C.~Clapson$^7$,
S.~Fairhurst$^4$,
D.~Grosjean$^5$,
G.~Guidi$^{6a,6b}$,
P.~Hello$^7$,
E.Katsavounidis$^3$,
M.~Knight$^1$,
A.~Lazzarini$^2$,
F.~Marion$^5$,
B.~Mours$^5$,
F.~Ricci$^9$,
A.~Vicer\'e$^{6a,6b}$,
M.Zanolin$^3$ - The joint LIGO/Virgo working group }

\address{ $^1$ Carleton College, Northfield MN 55057 USA}
\address{$^2$ LIGO-California Institute of Technology, Pasadena CA 91125
USA}
\address{$^3$ LIGO-Massachusetts Institute of Technology, Cambridge,
Massachusetts 02139 USA}
\address{$^4$ University of Wisconsin - Milwaukee, Milwaukee WI 53201 USA}
\address{$^5$ Laboratoire d'Annecy-le-Vieux de physique des particules,
Chemin de Bellevue, BP 110, 74941 Annecy-le-Vieux Cedex France }
\address{$^{6a}$ INFN - Sezione Firenze/Urbino
Via G.Sansone 1, I-50019 Sesto Fiorentino} 
\address{$^{6b}$ Universit\`a di Urbino,
Via S.Chiara 27, I-61029 Urbino Italia}
\address{$^7$ Laboratoire de l'Acc\'el\'erateur Lin\'eaire
(LAL), IN2P3/CNRS-Universit\'e de Paris-Sud, B.P. 34, 91898 Orsay Cedex
France}
\address{$^8$ INFN Sezione di Perugia and/or  Universit\`a di Perugia,
Via A. Pascoli, I-06123 Perugia Italia}
\address{$^9$ INFN, Sezione di Roma  and/or Universit\`a ``La Sapienza",
P.le A. Moro 2, I-00185, Roma Italia }

\eads{\mailto{zanolin@mit.edu} }


\begin{abstract}
We present a comparative study of 6 search methods for
gravitational wave bursts using simulated LIGO and Virgo noise data. 
The data's spectra were chosen to follow the design sensitivity of the 
two 4km LIGO interferometers and the 3km Virgo interferometer. 
The searches were applied on replicas of the data sets to which  
8 different signals were injected. 
Three figures of merit were employed in this analysis:
(a) Receiver Operator Characteristic curves,
(b) necessary signal to noise ratios for the searches to achieve 
50$\%$ and 90$\%$ efficiencies, and (c) variance and bias for the estimation of
the arrival time of a gravitational wave burst.
\end{abstract}
\submitto{\CQG}
\maketitle
\section{Introduction}
\label{sec:introduction}
Progress in the commissioning of the LIGO and Virgo detectors make possible 
in the near future the 
opportunity of a joint network analysis. To prepare the ground, gain a better 
understanding of each other's analysis and develop common procedures, the LIGO 
Scientific Collaboration and the Virgo experiment have agreed to pursue a joint 
search for burst and binary inspiral signals on simulated data \cite{ref:prop}. 
This paper describes the application of burst search algorithms 
on simulated data to which several classes of candidate signals waveforms were injected.
This framework provides the opportunity to (a) compare the performance of 
time domain and frequency domain methods (b) obtain a detector-independent measure 
of a method performance (c) test the limit of design sensitivity and (d) study separately 
the properties of the trigger generation from those of 
temporal coincidence modules between different interferometers.

Three hours of simulated data have been generated with a spectrum following the target design 
sensitivity of both the 4km LIGO interferometers, whose sampling frequency $f_s$ = 16384 Hz, 
and 3km Virgo interferometer, with  $f_s$ = $20000$ Hz (see the inspiral proceedings in this issue for 
more details \cite{ref:clapson}).

Three families of signals have been generated and injected onto the simulated noise with 
a mean Poisson rate equal to one every 60 seconds. Sine Gaussian (SG) and Gaussian (GA)  signals  
were chosen to represent the two general classes of short-lived gravitational wave bursts of 
narrow-band and broad-band character respectively. SG were chosen with central frequencies $f$ 
= $235$ Hz, to probe the best sensitivity region of the spectrum, and 820 Hz, to probe the 
higher frequency regime. Both of them were tested for quality factors $Q$ = $5$ and 15.
GA signals were chosen with duration equal to 1 and 4 milliseconds.
Dimmelmeier-Font-Mueller (DFM) supernovae core collapse signals \cite{ref:dimmel02} with parameters 
$a$ = $1, b$ = $2, g$ = $1$ and $a$ = $2, b$ = $4, g$ = $1$ 
were adopted as a more realistic model for gravitational wave bursts. DFM signals also allowed to probe 
the searches on waveforms with non-minimal time-frequency volume. 

The amount of energy in the signal was quantified in terms of the signal to noise ratio ($\rho$) typically 
used to characterize the performance of a an optimal filter
\begin{equation}
\rho=\sqrt{4\int_{0}^{\infty}\frac{|h(f)|^2}{\sigma(f)}df}  
\end{equation}
where $h(f)$ is the spectrum of the injected waveform and $\sigma(f)$ is the noise power spectral density.
The range of tested $\rho$ spanned between 2 and 10 because it roughly corresponds to the transition 
between not detecting any of the injected signals to detecting all of them for most of the searches involved.

The  methods considered can be divided into searches for excess power in the time frequency domain:   
\begin{itemize}
\item Q-transform (QT)\cite{ref:shour04}, a multiresolution time frequency search for excess power applied on data that are first whitened using a zero phase linear prediction. Equivalent to optimal matched filter for minimum uncertainty waveforms of unknown phase in the whitened data
\item S-transform (ST)\cite{ref:clapson2}, a search for statistically significant clusters of power in the 
time frequency map generated using a kernel composed of complex exponentials shaped by Gaussian profiles with 
width inversely proportional to frequency. The ST first whitened the data and also applied  on Virgo data a high pass filter and a line removal.
\item Power Filter (PF)\cite{ref:guidi}, a search on whitened data for excess power over different time intervals and set of frequencies 
\end{itemize}
and time domain searches: 
\begin{itemize}
\item Peak Correlator (PC)\cite{ref:arnaud99}, a search for peaks of Wiener filtered data with Gaussian templates. 
PC applies on Virgo data a high pass filter and a line removal filter.  
\item Mean Filter (MF)\cite{ref:pradier},  search for excess power in moving averages of whitened data over intervals containing from 10 to 200 samples. The MF whitens the data and applies to Virgo data a  high pass filter and a line removal.  
\item Adaptive Linear Filter ALF \cite{ref:arnaud03}, search for change in slope over moving windows of data over intervals containing from 
10 to 300 samples. ALF applies the same whitening high pass filtering and line removal as MF.
\end{itemize}
The performance of the searches have been investigated by computing (a) the Receiver Operator Characteristic curves 
(ROCs) for each method, injected waveform, interferometer and $\rho$ (b) computing the necessary $\rho$ for a search to 
reach an efficiency equal to 50$\%$ and 90$\%$ for a representative value of the False Alarm Rate (FAR) and
(c) assessing the accuracy in the estimate of the arrival peak time of a signal.

In section 2 we describe the analysis pipeline. In section 3 we discuss a selection of the figures of merit 
generated to compare the searches. In section 4 we summarize the findings of this project and address future 
directions of the investigation.   
\section{Analysis Pipeline}
Calibrated strain data corresponding to noise and signal from both instruments were made available in the frame 
format to both the collaborations. Each method has then been applied to noise plus signals for FARs in the range 
between $10^{-4}$Hz, which corresponds to roughly one false trigger in three hours,  and 0.1Hz, which allowed to 
have more noise triggers and study the statistical distribution of durations and temporal separation of the triggers. 
The performance of the searches at higher single interferometer FARs is also relevant for network studies where consistency 
criteria (for example on the time of arrival and amplitude) can be applied on the triggers giving a lower FAR for 
consistent events.

Each method computed a start time, a peak time and an end time for each of the events, as well as the efficiencies 
in the chosen range of FARs. The searches adopted different procedures to associate triggers  with injections and 
therefore compute the efficiencies. However methods that use matching windows to identify detected injections with 
triggres can freely choose the size of the matching window provided that it is (I) larger than the search time 
resolution and (II) the products of the matching window for the false alarm rate and the injection rate are both much 
smaller than one. In this analysis: (a) the PF associated to each injected waveform a time interval starting 
at the first non zero sample and finishing at the last non zero sample. A detection was claimed when the trigger 
interval, as computed by the PF, overlapped with the event interval. (b) For the QT, the interval internally 
associated to a trigger needed to overlap with a 0.2 seconds interval centered at the the injection peak time. 
(c) For MF, ALF, PC and S the time associated with a trigger needed to be closer to the peak time of the injection
than 20 ms for ALF, MF and S, 50 ms for PC.\\
As we will see in section 3, were we discuss the accuracy of the peak time estimation, the constraint (I) was respected. 
The constraint (II) was also respected. Infact the largest FAR that we study is 0.1Hz
and the largest coincidence window is 0.05 seconds while the injection rate is 1/60 Hz.
This makes the two products of point (II) both smaller than at least 0.01. 
\section{Data Analysis}
\label{sec:data_analysis}
A key element in comparing search methods are ROCs because they allow one to confront the efficiencies in a 
range of FARs of interest. For example in a single-interferometer experiment the FAR could be chosen by the request 
of having only less than one noise event over the whole acquisition time. More generally, the largest tolerable 
single interferometer FAR will depend on the observation time, number of interferometers and consistency requirements between triggers 
generated from the data belonging to different interferometers. In this perspective as informative figures of merit we 
first present examples of ROC curves and the necessary $\rho$ to reach 50$\%$ and 90$\%$ detection efficiencies 
for FAR = 0.01Hz. Of all the possible parameters on which consistency constraints can be applied, in this 
study we focused on the trigger peak time, which should be compatible with the travel time between interferometers and 
the methods errors in the peak time estimation. In the summary tables 1 to 4 some results corresponding to the PC 
and ST running on SG are not available. They reflect waveforms where the efficiency of the PC was not expected to 
be significant or the ST had not completed their analysis. We plan to include these results in future publications.

We present representative ROC curves for SG235Q5 and DFMa1b2g1, two values of $\rho$ (10 and 5) on Virgo data in figures 1 and 2,
while the ROC curves for SG820Q15 and GA1d0, two values of $\rho$ (10 and 5), on LIGO data in figures 3 and 4. ROCs for the same 
waveform and different interferometers are not shown since the performance of the methods gave results fairly similar over the two 
noise spectra.  Time domain methods were observed to perform better on GA pulses and DFM  supernovae core collapse 
while time-frequency domain methods had a better performance on the SG signals. The two families of curves 
corresponding to the two values of $\rho$ of the injected signals also illustrate how the algorithm  
performance degrades with decreasing $\rho$. SG results follow intuition since the time frequency 
tailoring of the QT and ST is close to a matched filter for minimum uncertainty waveforms of unknown phase in the 
whitened data. A similar line of thought applies to PC which is an optimal filter for GA pulses. 
\begin{figure}[htbp]
  \begin{center}
  \includegraphics[height = 9cm,width=17cm]{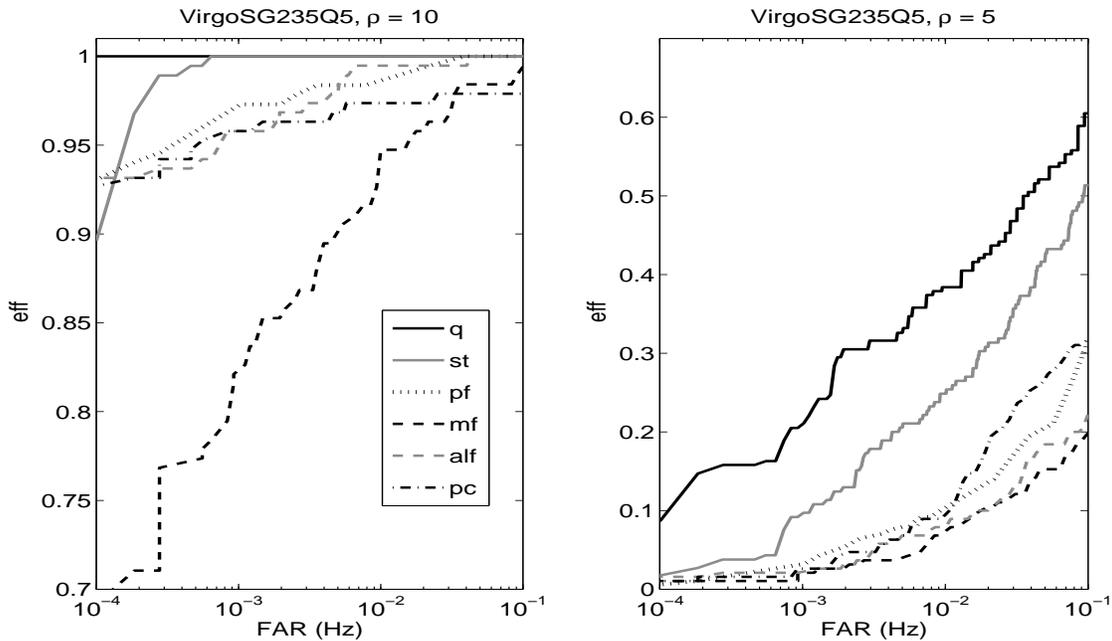}
  \end{center}
  \caption{
  Virgo  ROCs for sg235q5 and $\rho$ = 10 (left) and 5 (right).
 MF is in black dashed, ALF in gray dashed,  PC in black dashdot, PF in black dotted, ST in solid gray and QT in solid black}
  \label{fig:algorithm}
\end{figure}
\begin{figure}[htbp]
  \begin{center}
  \includegraphics[ height=9cm,width=17cm]{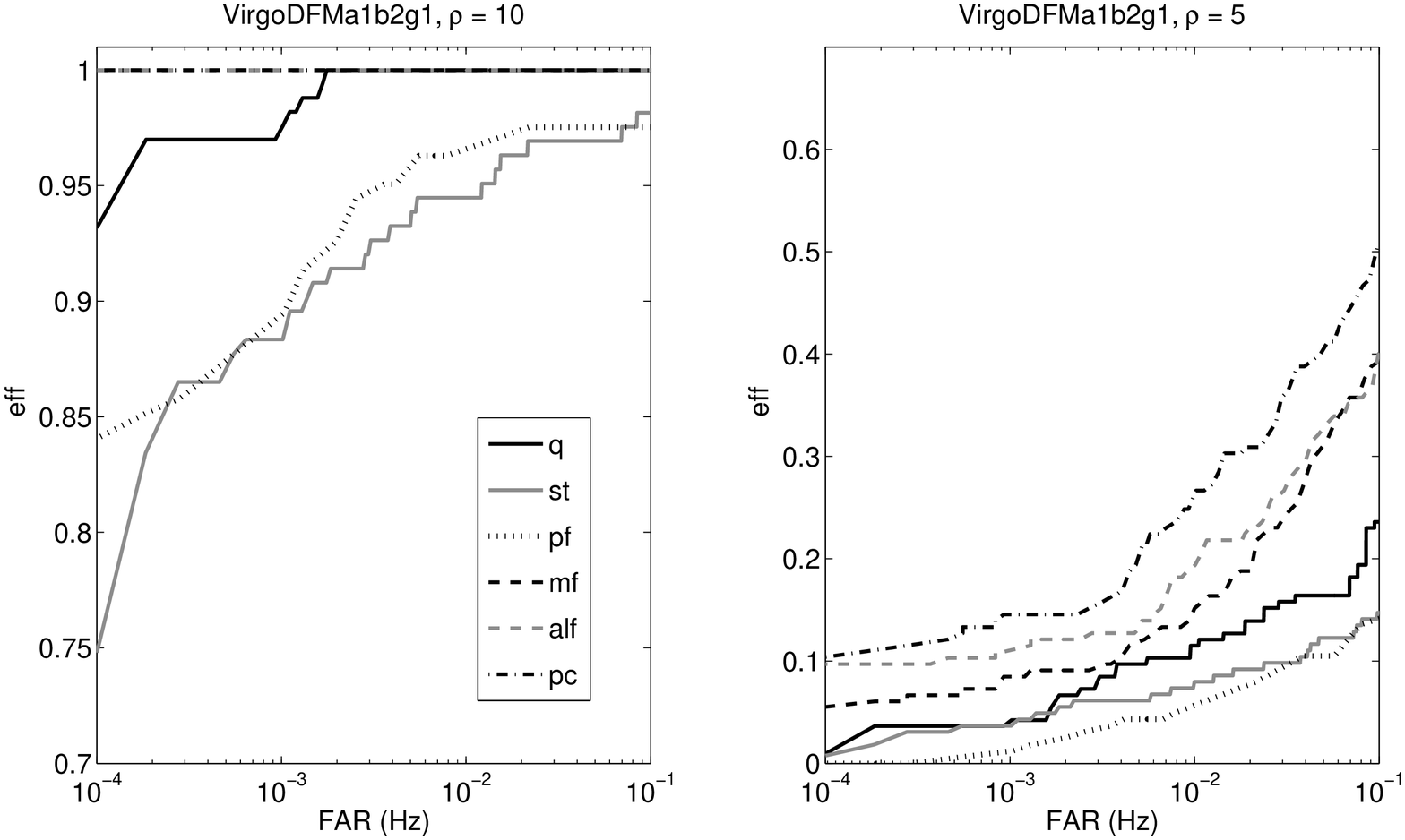}
  \end{center}
  \caption{
  Virgo ROCs for dfma1b2g1 and $\rho$ = 10 (left) and 5 (right).
MF is in black dashed, ALF in gray dashed,  PC in black dashdot, PF in black dotted, ST in solid gray and QT in solid black}
  \label{fig:algorithm}
\end{figure} 
 \begin{figure}[htbp]
  \begin{center}
  \includegraphics[height=9cm,width=17cm]{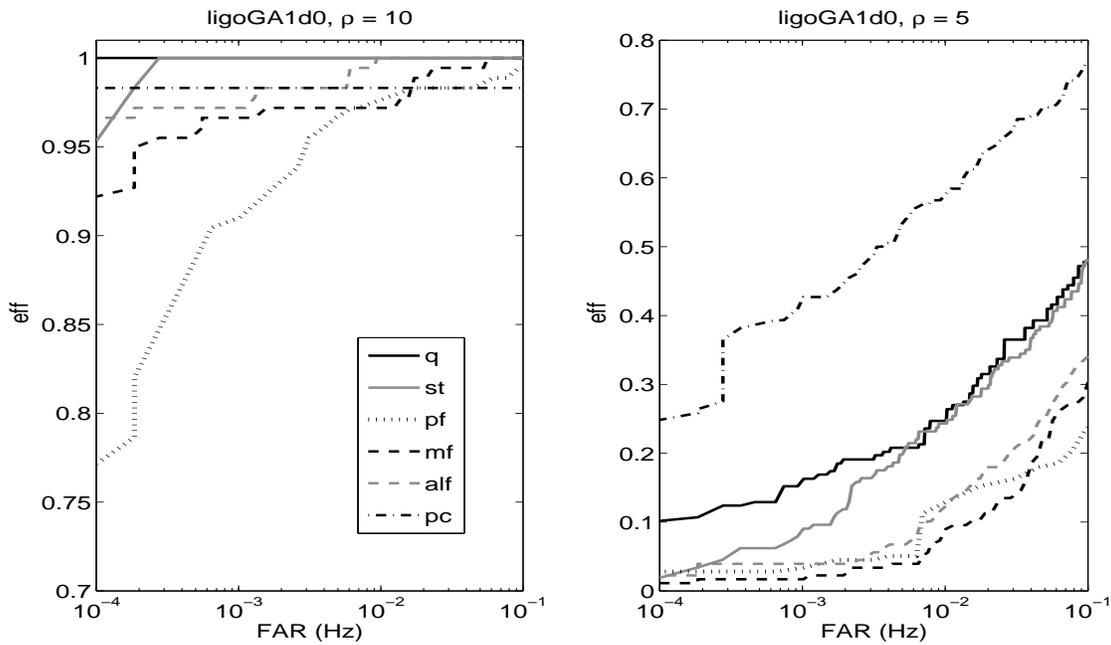}
  \end{center}
  \caption{
  LIGO ROCs for a 1 millisecond GA signal and $\rho$ = 10 (left) and 5 (right).
 MF is in black dashed, ALF in gray dashed,  PC in black dashdot, PF in black dotted, ST in solid gray and QT in solid black.}
  \label{fig:algorithm}
\end{figure}
\begin{figure}[htbp]
  \begin{center}
  \includegraphics[height=9cm,width=17cm]{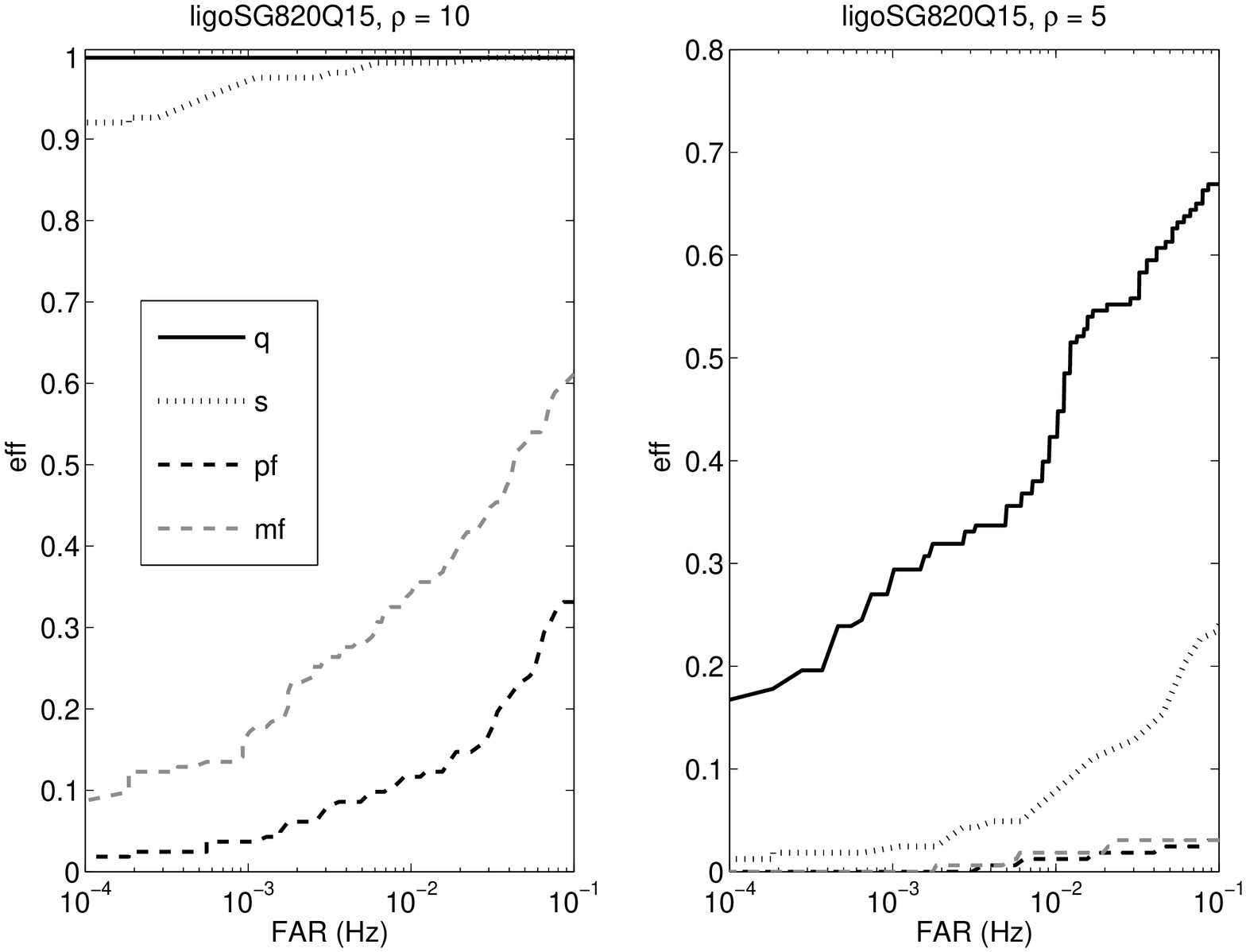}
  \end{center}
  \caption{
  LIGO ROCs for a SG of central frequency equal to 829 Hz and Q = 5.
 MF in gray dashed,  PF in black dashed, ST in black dotted and QT in solid black.}
  \label{fig:algorithm}
\end{figure}
The performance of search is often quantified through the conditions for which, at the given FAR, a search 
achieves 50 $\%$ and 90 $\%$ efficiency. We present these conditions here in terms of $\rho$ since it is the 
parameter that quantifies the detectability of a signal in optimum filtering. In table 1 and 2 we present these 
$\rho$ for a FAR = 0.01Hz which would correspond to less than a false event in 3 hours, if a temporal 
coincidence between Virgo and one LIGO  interferometers would be imposed. This is reasonable because if the noise triggers 
are generated randomly with a Poisson distribution, as is natural to expect for a stationary interferometer, 
the rate of coincidence can be estimated. The result is roughly given by the product of the two rates
by twice the length of the temporal coincidence window, which can be bounded with 50 milliseconds for the searches 
involved in this study. Explicitly, the values of the necessary $\rho$ are computed in two steps:
(a) the efficiencies for a given FAR and different $\rho$ are extracted from the ROC curves and (b) a numerical 
interpolation is performed with a least square fit that minimizes the free parameters $\alpha, \beta, \gamma$ of the 
asymmetric sigmoid 
\begin{equation}
E(\rho)=\frac{1}{1+(\frac{\rho}{ \gamma}) ^{\alpha(1+\beta tanh(  \frac{\rho}{\gamma}))  }}.
\end{equation}
where $\gamma$ corresponds to the value of $\rho$ for which the efficiency    $E$ = 0,5 
$\beta$ is the parameter that describes the asymmetry of the sigmoid (and takes values between -1 and +1)
and $\alpha$ describes the slope.
\begin{table}
\caption{necessary $\rho$ for .5 and .9 efficiencies at FAR = 0.01Hz for LIGO data }
\begin{tabular}{|r|r|r|r|r|r|r|r|r|r|r|r|r|}\hline
LIGO eff & mf .5   &  .9 & alf .5 &  .9 & pc .5 &  .9 & pf.5 & .9  & s .5  &  .9 &  q .5 & .9  \\  
\hline
sg235q5   &  7.5  & 10.2  &  6.4  &  8.2  &  6.1  &  8.4  &  6.8  &  8.8  &  5.2  &  6.3  &  5.1 &   6.4 \\
\hline
sg235q15  &  10.9  & 15.5  &  9.2  & 12.6  & NA  &  NA  &  6.3  &  8.5  &  5.4  &  6.7  &  5.1 &   6.5 \\
\hline
sg820q5   &  8.7  & 11.7  &  7.4  &  9.5  & NA  &  NA  &  7.6  &  9.9  &  NA  &  NA  &  5.2 &   6.6 \\
\hline
sg820q15  &  15.4  & 24.4  & 11.1  & 15.4  & NA &  NA  &  6.8  &  8.5  & NA  &  NA  &  5.1 &   6.3 \\
\hline
ga1d0   &  6.6  &  8.6  &  6.3  &  7.9  &  4.9  &  6.2  &  7.1  &  9.1  &  5.5  &  6.8  &  5.5 &   6.7 \\
\hline
ga4d0  &   7.5  &  9.6  &  6.4  &  8.1  &  4.9  &  6.0  &  6.8  &  8.6  &  5.9  &  7.4  &  5.6 &   7.0 \\
\hline
dfma1b2g1  &   7.3  &  9.3  &  6.7  &  8.6  &  5.1  &  6.4  &  7.8  & 10.1  &  6.9  &  8.4  &  6.5 &   8.2 \\
\hline
dfma2b4g1  &   6.9  &  9.2  &  6.6  &  8.6  &  5.8  &  7.5  &  8.0  & 10.3  &  6.6  &  8.2  &  6.4 &   8.0 \\
\hline
\end{tabular}
\end{table} 
\begin{table}
\caption{necessary $\rho$ for .5 and .9 efficiencies at FAR = 0.01Hz for Virgo data }
\begin{tabular}{|r|r|r|r|r|r|r|r|r|r|r|r|r|}\hline 
 Virgo eff  & mf .5   &  .9 & alf .5 & .9 & pc .5 &  .9 & pf.5 &  .9  & s .5  &  .9 &  q .5 & .9  \\   
\hline
sg235q5   &  7.5  &  9.9  &  6.9  &  8.7  &  6.6  &  8.7  &  6.5  &  8.4  &  5.7  &  7.2  &  5.2 &   6.4 \\
\hline
sg235q15  &  11.1  & 16.5  &  9.6  & 12.6  & NA  &  NA  &  6.4  &  8.4  &  5.6  &  7.0  &  5.1 &   6.6 \\
\hline
sg820q5  &   8.1  & 10.6  &  6.8  &  8.4  &  NA  &  NA  &  7.6  &  9.6  & NA  &  NA  &  5.1 &   6.5 \\
\hline
sg820q15  &  19.1  & 23.8  &  9.9  & 13.6  & NA  &  NA  &  7.3  &  9.4  & NA  &  NA  &  5.3 &   6.8 \\
\hline
ga1d0  &   5.8  &  7.6  &  5.6  &  7.1  &  4.9  &  6.0  &  7.2  &  9.1  &  5.6  &  7.0  &  5.9 &   7.5 \\
\hline
ga4d0  &   6.2  &  7.8  &  5.3  &  6.7  &  5.2  &  6.1  &  8.7  & 11.7  &  5.4  &  6.8  &  7.0 &   9.1 \\
\hline
dfma1b2g1  &   6.1  &  7.7  &  5.9  &  7.5  &  5.6  &  6.9  &  7.7  &  9.4  &  5.7  &  7.2  &  6.6 &   8.2 \\
\hline
dfma2b4g1  &   6.3  &  8.1  &  6.0  &  7.6  &  5.8  &  7.5  &  8.3  & 10.9  &  5.5  &  6.9  &  6.8 &   8.8 \\
\hline
\end{tabular}
\end{table}
It is interesting to notice from table 1 and 2 that ST and QT which are both 
based on a multiresolution decomposition of the time frequency plane, have similar necessary $\rho$ to obtain the 
50 $\%$ and 90 $\%$ efficiencies. Time domain methods similarly to the ROC curves require higher $\rho$ than for SG that for
GA and DFM waveforms.

The accuracy of the estimate of the arrival time of a pulse
determines how strict can be time consistency cuts typically involved in the post processing 
of the triggers and the angular resolution of a network of detectors. 
Explicitly, in order to identify the direction of arrival of a gravitational wave,
the error in the peak time estimation needs to be much smaller than the travel times 
between LIGO interferometers (~10 milliseconds) and between Virgo and one of the LIGO interferometers
(~20 milliseconds). A more quantitative analysis of the angular resolution is the goal of an ongoing project
of this collaboration. 
 
We present this information here in terms of the standard deviation and bias, both in milliseconds,
\begin{equation}
b=\frac{1}{N} \sum_{i=1}^{N} (\widehat{t}_i-t_i),\,\,\,\,  std=\sqrt{ \frac{1}{N-1} \sum_{i=1}^{N} (  (\widehat{t}_i-t_i)-b   )^2  }
\end{equation}
of the estimate of the peak arrival time $\widehat{t}_i$ with respect to the peak time of the injections $t_i$
for $\rho$ = 10 and FAR = 0.1Hz (which provides the largest statistics among the FARs we studied).
\begin{table}
\caption{std in ms for peak time estimation $\rho$ = 10 FAR = 0.1Hz}
\begin{center}
\begin{tabular}{|r|r|r|r|r|r|r|}\hline  
standard-d & mf &  alf & pc&  pf&  s  & q\\
\hline     
sg235q5 ligo  &  1.4  &  1.2  &  1.2 &  21.7  &  0.7  &  0.7\\
\hline
sg235q5 virgo  &  1.5  &  1.3  &  1.3 &  13.8  &  0.8  &  0.9\\
\hline 
sg235q15 ligo  &  4.1  &  4.2  &  NA    &  15.2  &  1.7  &  1.9\\
\hline
sg235q15 virgo  &  4.9  &  5.2  &  NA    &  11.7  &  2.0  &  2.5\\
\hline 
sg820q5 ligo  &  0.4  &  0.4  &  NA    &  15.9  &   NA    &  0.2\\
\hline
sg820q5 virgo  &  0.4  &  0.3  &  NA    &  16.9  &  NA     &  0.2\\
\hline 
sg820q15 ligo  &  1.3  &  1.3  &  NA    &  19.1  &  NA     &  0.6\\
\hline
sg820q15 virgo  &  8.4  &  3.3  &  NA    &  14.5  &  NA     &  0.7\\
\hline 
ga1d0 ligo  &  1.2  &  1.6  &  0.1 &  15.5  &  0.7  &  0.7\\
\hline
ga1d0 virgo  &  1.2  &  0.6  &  0.1 &  16.4  &  0.8  &  0.8\\
\hline 
ga4d0 ligo  &  1.6  &  3.5  &  0.3 &  15.5  &  5.8  &  1.5\\
\hline
ga4d0 virgo  &  2.5  &  1.6  &  0.3 &  12.3  &  1.4  &  1.9\\
\hline 
dfma1b2g1 ligo  &  0.3  &  1.0  &   0   &  23.0  &  1.4  &  0.7\\
\hline
dfma1b2g1 virgo  &  0.3  &   0    &   0   &  15.2  &  0.4  &  0.2\\
\hline 
dfma2b4g1 ligo  &  1.7  &  2.8  &  1.0 &  18.1  &  2.6  &  2.2\\
\hline
dfma2b4g1 virgo  &  2.5  &  0.3  &  0.2 &  18.1  &  2.5  &  1.4\\
\hline
\end{tabular}
\end{center}
\end{table}
\begin{table}
\caption{bias in ms for peak time estimation $\rho$ = 10 FAR = 0.1Hz}
\begin{center}
\begin{tabular}{|r|r|r|r|r|r|r|}\hline  
bias & mf &  alf & pc &  pf&  s  & q\\
\hline 
sg235q5 ligo  &  1.8  &  0.2  &  0.2 & -5.0  &  1.7  & -0.1\\
\hline
sg235q5 virgo  &  0.7  & -0.5  & -0.5 & -3.9  &  0.4  & -0.1\\
\hline 
sg235q15 ligo  &  1.9  &  0.3  &  NA    & -5.8  &  1.5  & -0.2\\
\hline
sg235q15 virgo  &  0.6  & -0.1  &   NA   & -9.5  &  0.5  &   0  \\
\hline 
sg820q5 ligo  &  0.1  & -0.4  &   NA   & -7.2  &   NA    &  0   \\
\hline
sg820q5 virgo  &  0.1  & -0.4  &  NA    & -5.1  &   NA    &  0   \\
\hline 
sg820q15 ligo  &  0.3  & -0.4  &  NA    & -4.8  &   NA    & -0.1 \\
\hline
sg820q15 virgo  &  1.1  &    0   &  NA    & -5.1  &   NA    &  0  \\
\hline 
ga1d0 ligo  &  1.2  &  3.3  &   0   & -0.4  &  3.7  &  0   \\
\hline
ga1d0 virgo  & -0.7  &  2.3  &   0   & 0.7  &  1.6  & -0.1\\
\hline 
ga4d0 ligo  &  3.3  &  8.0  &   0   & 4.4  &  2.9  &   0  \\
\hline
ga4d0 virgo  & -0.2  &  3.3  &   0   & -3.6  &   0   & -0.2\\
\hline 
dfma1b2g1 ligo  & -0.2  &  0.6  & -0.1 & -8.6  &  1.5  &  0.2\\
\hline
dfma1b2g1 virgo  & -0.4  &    0   & -0.1 & -6.1  &   0    & -0.1\\
\hline 
dfma2b4g1 ligo  &  2.4  &  4.4  & -1.0 & 4.6  &  4.6  & -0.9\\
\hline
dfma2b4g1 virgo  & -2.8  &  2.0  & -1.3 & -1.0  &  2.6  & -1.6\\
\hline 
\end{tabular}
\end{center}
\end{table}
The accuracy of the searches appears to be similar between LIGO and Virgo data.
Most of the methods can be used for a directional search on most the waveforms since the 
time accuracy is typically smaller than the travel time between the interferometers. 
It is also worth notice that the feasibility of directional searches 
rapidly degrades with decreasing $\rho$ and that, if a method shows a systematic negative bias, 
this can be usually easily corrected.



\section{Conclusions}
The main motivation of this work was to prepare the ground to 
joint searches of gravitational burst signals between the LIGO and Virgo interferometers.
In particular in this work we built a joint framework where 
we learned how to exchange interferometer data, trigger files and analyze each other's data. 
We also gained a deeper understanding of the operational properties of our burst 
searches. This analysis showed that the performance of time domain and frequency domain methods 
have different strengths on different kinds of waveforms. The performance of the searches was 
fairly stable shifting between LIGO and Virgo data. The study of the 
arrival time estimation accuracy showed that most of the methods can be employed in directional 
searches if the $\rho$ is high enough and gave an indication of how stringent can be temporal 
consistency cuts between different interferometers. If a combination of the methods had to be
applied on real data it would need to be studied as a search in itself.
In particular more extensive studies would be necessary to verify if the ROC curve of the combination 
of methods stands above the ROCs of the single searches.
The next steps of this work are to extend this comparisons on directional searches
to be performed on simulated data corresponding to an network of 3 interferometers,
investigate more the usefulness of combination of methods and explore the use of consistency 
criteria other than the arrival time. It is also in the plans to expand the number of searches
involved in the study. 
\vspace{0.1in}
\noindent
\\
LIGO Laboratory and the LIGO Scientific Collaboration gratefully
 acknowledge the support of the United States National Science Foundation for the construction and operation of the LIGO Laboratory and for the support of this research.


\end{document}